\documentclass[aps,prb,twocolumn,reprint,preprintnumbers,amsmath,amssymb,floatfix]{revtex4-1}
\usepackage{latexsym}
\usepackage{dcolumn}
\usepackage[dvips]{graphicx}
\usepackage{amssymb}
\usepackage{graphics}
\usepackage{float}
\usepackage{amsmath}
\usepackage{epsf}
\newcommand{\beq}{\begin{equation}}
\newcommand{\eeq}{\end{equation}}
\usepackage{bm}

\newcommand{\vk}{{\bf k}}
\newcommand{\vq}{{\bf q}}
\newcommand{\eps}    {\epsilon}
\newcommand{\ceq}[1] {(\ref{#1})}
\newcommand{\rr}     {{\bf r}}

\newcommand{\kk}     {{\bf k}}
\DeclareMathOperator {\sgn}{sgn}

\newcommand{\nimp}   {n_{\rm imp}}

\newcommand{\br}     {\mathbf{r}}
\newcommand{\nrms}   {n_{\rm rms}}
\newcommand{\nav}   {\langle n\rangle}

\newcommand{\sigmaena}{\sigma_e^{(na)}}
\newcommand{\sigmahna}{\sigma_h^{(na)}}
\newcommand{\sigmaea}{\sigma_e^{(a)}}
\newcommand{\sigmaha}{\sigma_h^{(a)}}
\newcommand{\smin}{\sigma_{\rm min}}

\graphicspath{{../figures/}}
\begin{document}
\title{Two-dimensional electronic transport on the surface of 3D topological insulators}
\author{Qiuzi Li$^1$, E. Rossi$^2$, and S. Das Sarma$^1$}
\affiliation{$^1$Condensed Matter Theory Center, Department of Physics, University of Maryland, College Park, Maryland 20742, USA\\
             $^2$Department of Physics, College of William and Mary, Williamsburg, VA 23187, USA}
\date{\today}

\begin{abstract}

We present a theoretical approach to describe the 2D transport properties of the
surfaces of three dimensional topological insulators (3DTIs) including disorder and phonon scattering effects.
The method that we present is able to take into account the effects
of the strong disorder-induced carrier density
inhomogeneities that characterize the ground state of the surface
of 3DTIs, especially at low doping, as recently shown experimentally.
Due to the inhomogeneous nature of the carrier density landscape, standard
theoretical techniques based on ensemble averaging over disorder assuming a spatially uniform average carrier density are inadequate. Moreover the presence of strong
spatial potential and density fluctuations greatly enhance the effect of thermally activated
processes on the transport properties. The theory presented is able
to take into account all the effects due to the disorder-induced inhomogeneities, momentum scattering by disorder, 
and the effect of electron-phonon scattering
processes. As a result the developed theory is able to accurately describe
the transport properties of the surfaces of 3DTIs both at zero
and finite temperature.

\end{abstract}

\maketitle

\section{Introduction}
\label{sec:intro}

In strong three dimensional topological insulators (3DTIs) the nontrivial topology of the
bulk energy bands \cite{HasanKane_RMP10, FKM_PRL07,Culcer_PhysicaE2011}
enforces the essential existence of 2D metallic surface states that can be well described at low energies as massless
Dirac fermions. The valence and conduction band of the surface states touch at isolated
points, the Dirac points, whereas the bulk states are gapped.
Angle resolved photo-emission spectroscopy experiments have confirmed the existence of
the Dirac like surface states in Bi$_{1-x}$Sb$_x$\cite{Hsieh_Nature08}, Bi$_2$Se$_3$\cite{Hasan_NPH09,Hsieh_Nature09},
Bi$_2$Te$_3$\cite{chen_science_2009,Hsieh_PRL09} and Sb$_2$Te$_3$\cite{Hsieh_PRL09}.
Electronically, the surface of a 3DTI is very analogous to graphene
\cite{novoselov2004,dassarma2011}
in which the fermionic excitations are also well described, at low energies,
as massless Dirac fermions. There are two fundamental differences between graphene
and the surfaces of 3DTIs:
({\bf i})  In graphene the chiral nature is due to the locking of the momentum direction with the electron pseudospin, associated
           with the sublattice degree of freedom, instead of the real spin as in the surface of strong 3DTIs;
({\bf ii}) In strong 3DTIs the number of Dirac points is odd whereas in graphene is even.
These differences make graphene and the surfaces of strong 3DTIs fundamentally different.
However, the fact that graphene and the surfaces
of strong 3DTIs are both two-dimensional electronic systems and have a very similar band structure
suggests that these two systems might have similar charge-transport properties.
As we show in this work this is only partially correct.

Two important aspects of TI surface transport need to be mentioned (in the context of our comprehensive theoretical work to be presented in this paper) so as to avoid any confusion about our goal and scope.  First, 2D TI transport occurs on the surface of 3D TI materials, and theoretically the bulk 3D states should be insulating with no metallic contribution to the conductivity.  Experimentally, however, this situation of 2D metallic transport on a bulk 3D insulator has not yet been achieved since the bulk, instead of being a band insulator, seems to have a lot of free carriers which contribute (indeed, often dominate) the measured conductivity\cite{Butch_PRB10}.  We completely ignore the complications of the bulk conduction and the bulk carriers in our theory concentrating entirely on purely the 2D surface conductivity (as a function of density and temperature) assuming the bulk to be an insulator as the theory implies it should be.  Recent experimental advances in materials preparation and thin film device fabrication have made it possible to see the expected pure 2D surface conduction with little contamination from the bulk states, and thus, our work is relevant to an increasing body of recent data\cite{Dohun_arXiv11,Hong_arXiv11,Kong_Natnano,Steinberg_PRB11}.  In any case, the subject is interesting only because of the 2D metallic surface topological states, and therefore, we focus entirely on this issue.  Second,  the actual energy dispersion of the 3DTI surface states appears to follow the linear Dirac-like spectrum only at rather low energies, particularly for the hole states (i.e. the valence band) with strong nonlinearity becoming apparent at higher energies.  This nonlinearity (e.g. a quadratic correction to the linear dispersion) is nonuniversal and strongly materials-dependent whereas the theoretical Dirac-behavior at low energies is universal (with different Fermi velocities characterizing different TI materials).  Because the focus of our work is on the universal behavior arising from density inhomogeneity (which is much more important at low Fermi energies, i.e. low 2D densities), which should not depend much on the details of the band structure at higher energies, we ignore the higher energy parabolicity of the 2D surface bands.  We emphasize that these two approximations, neglect of bulk conduction and surface band dispersion correction at higher energy, imply that our theory should not be construed as a quantitative theory for any particular TI material, but as a qualitative guide for the universal features of 3DTI surface transport behavior.  Detailed quantitative comparison between theory and experiments is further complicated by our lack of knowledge of the precise parameters for TI systems any way (e.g. the Fermi velocity, the phonon parameters, the nature of disorder, etc.), and therefore, our theory provides the zeroth order qualitative theory for 3DTI surface transport which should apply to all TI materials.

The surfaces of the newly discovered strong 3DTIs are of great fundamental interest\cite{Wilczek_PRL87,qxl_PRB08,Wilczek_Nat09,rli_natphy10}
and in addition have the potential to be used in disruptive novel technologies such as
topological quantum information processing\cite{FuKane_PRL08,FuKane_PRB09,JMoore_Nature10}.
To be able to use the surfaces of 3DTIs to study novel fundamental phenomena and in novel technologies
it is essential to understand their electron transport properties and in particular
to understand the main factors limiting their electron mobility.
Recently, experiments on thin films of strong 3DTIs
\cite{Dohun_arXiv11,Hong_arXiv11,Kong_Natnano,Steinberg_PRB11},
by enhancing the  surface-to-volume ratio\cite{LiuMinhao_PRB11},
have been able to greatly reduce the bulk contribution transport and obtain the intrinsic 2D transport properties of the
surfaces of strong 3DTIs.
Previous theoretical works on the transport in the surface of strong 3DTIs
\cite{Culcer_TIPRB10,Adam_arXiv12}
have used simplified models and have mostly focused
on the zero temperature limit.

In this work we present a comprehensive transport theory for the surfaces of 3DTIs
valid both at zero temperature and at finite temperatures.
As in graphene \cite{Martin_NP08,YZhang_NP09},
one of the main difficulties in developing a transport theory
for the surfaces of 3DTIs is the presence, especially at low doping,
of strong carrier density inhomogeneities
\cite{Beidenkopf_NPH11} induced by disorder.
The presence of strong spatial fluctuations makes the standard
theoretical approaches, that rely on the homogeneous nature of the ground state,
inadequate. Moreover, because of the inhomogeneities, at finite temperatures
the contribution of thermally activated carriers to transport can be
very important and difficult to quantify. Thus, disorder-induced density inhomogeneity by itself could introduce considerable insulating-like activated transport behavior in the nominally metallic 3DTI surface conduction.
The transport theory that we present takes into account both the effects
of quenched disorder and electron-phonon scattering processes.
In particular our theory is able to take into account the effects of the strong
disorder-induced inhomogeneities both at zero and finite temperature.

To characterize the inhomogeneous ground state we generalize to the case
of 3DTIs' surfaces the Thomas-Fermi-Dirac-Theory (TFDT) first developed
to study graphene \cite{rossi2008,dassarma2011}.
Combining the TFDT results and the Boltzmann theory we develop and validate
the effective medium theory (EMT) to obtain the transport properties
at zero temperature. We then develop an effective 2-fluid transport
theory that we validate at zero temperature by comparing its results
to the ones obtained using the EMT.
The great advantage of the 2-fluid theory is that it allows to
readily obtain the transport properties at finite temperature
including all the temperature dependent effects:
electron-phonon scattering, thermal activation, changes with
temperature of the screening properties, and thermal broadening
of the Fermi surface.
Given the experimental evidence \cite{Dohun_arXiv11,Hatch_PRB11}
that in current 3DTIs charged impurities
are the dominant source of disorder we have applied the theory
to the case in which the quenched disorder is due to random
charges placed in the vicinity of the surface of the 3DTI. We provide both a detailed analytical theory and comprehensive numerical results for the density and temperature dependent 3DTI surface transport properties in the presence of density inhomogeneity, scattering by random charged impurities, and phonon scattering.

%Our results show important qualitative difference between the transport
%properties of the surfaces of 3DTIs and graphene. In particular we find
%that the conductivity depends much more strongly on the temperature
%in 3DTI than in graphene due to the fact that in 3DTIs the electron-phonon
%scattering processes are much more important than in graphene.

In section~\ref{sec:theory} we present the theoretical approach that
we have developed to describe the transport on the surfaces of 3DTIs
taking into account both quenched disorder and electron-phonon scattering processes,
in particular when the quenched disorder is due to charged impurities.
In section~\ref{sec:results:tfdt} we present our results for the characterization
of the disorder-induced carrier density inhomogeneities obtained using
the TFDT. In section~\ref{sec:zeroTcond} we present our results for
the conductivity at zero temperature and finally in section~\ref{sec:finiteTcond}
we present our results for the conductivity at finite temperature.
Section~\ref{sec:conclusions} briefly summarizes our findings
and the differences between the 2D transport properties of the surfaces
of 3DTIs and single layer graphene.

\section{Theoretical approach}
\label{sec:theory}

To study the electronic transport on the surface of strong 3DTIs we use the Boltzmann theory.
From the Boltzmann theory, within the ``relaxation time approximation'', the electronic
conductivity $\sigma$ is given by the following equation:
\begin{equation}
 \sigma(n,T) = \frac{e^2}{2} \int d\epsilon D(\epsilon) v_F^2 \tau_{tot} (\epsilon,T)\left(-\frac{\partial f(\epsilon)}{\partial \epsilon}\right)
\label{eq:lcond}
\end{equation}
where $e$ is the electron charge, $D(\eps)$ the density of states at energy $\eps$, $v_F$ the Fermi velocity, $f(\eps) = 1/(1+ e^{\beta (\eps-\mu)})$ the Fermi-Dirac distribution,
and $\tau_{tot}$ the total transport mean free time due to the electron scattering off quenched disorder and phonons.
Assuming independent scattering from disorder and phonons (we mention that this is not equivalent to assuming the Matthiessen's rule which assumes that the resistivity due to independent mechanisms can be added and is invalid for our system) we have
\begin{equation}
 \frac{1}{\tau_{tot}} = \frac{1}{\tau_{dis}}+\frac{1}{\tau_{ph}}
\end{equation}
where $\tau_{dis}$ is the transport mean free time due to electrons scattering off quenched disorder and $\tau_{ph}$ is the
transport mean free time due to electron-phonon scattering processes.
One thing that we must emphasize is that Eq.~\ceq{eq:lcond} is valid as long as the system is homogeneous (i.e. spatial density fluctuations effects are small enough so that the average density $n$ is a meaningful quantity, an approximation which would break down for low $n$).
%Often in 2D systems the disorder potential induces strong inhomogeneities; in this situation Eq.~\ceq{eq:lcond}
%in general is not valid. However, a local form of Eq.~\ceq{eq:lcond} can be used as long as the characteristic
%length scale of the inhomogeneities is much larger than the mean free path $\ell= v_F\tau_{tot}$.

The energy dependent scattering time $\tau_{dis}(\epsilon)$ due to quenched disorder is given by
\begin{eqnarray}
 \dfrac{\hbar}{\tau_{dis}(\epsilon_{p{\bf k}})}= 2\pi n_{dis} \int \frac{d^2 k'}{(2\pi)^2}|\langle V_{p{\bf k},p{\bf k'}}\rangle|^2 g(\theta_{\bf kk'})\nonumber \\
 \times \left[1-\cos\theta_{\bf kk'}\right]\delta(\epsilon_{p\mathbf{k'}}-\epsilon_{p\mathbf{k}})
 \label{eq:mscatt}
\end{eqnarray}
where $\epsilon_{p\mathbf{k}} = p \hbar v_F |\mathbf{k}|$ is the energy of a quasiparticle
with $p=\pm 1$ and momentum ${\bf k}$, $n_{dis}$ is the 2D density of impurities,
%impurity centers giving rise to the random disorder
%potential\cite{DasEnrico_PRB10,qzli_MLGPRB11},
$\langle V_{p{\bf k},p{\bf k'}}\rangle$ is the matrix element of the
scattering potential,
$g(\theta_{\bf kk'})=\left[1+\cos\theta_{\bf kk'}\right]/2$ is the TI chiral matrix element factor
arising from the wavefunction overlap between states with momentum $\kk$
and momentum $\kk'$ with $\theta_{\bf kk'}$ the angle between $\kk$ and $\kk'$.
In Eq.~\ceq{eq:mscatt}, to minimize the number of parameters entering the theory we have assumed
that the impurities are randomly
distributed in a 2D plane located at an effective distance $d$ from the surface of the 3DTI. It is straightforward to include in the theory a more complex three-dimensional distribution of quenched impurities, but given the lack of experimental information about the distribution of unintentional and unknown quenched impurity disorder in the system, it is theoretically more meaningful to use a minimal model with just two unknown parameters $n_{dis}$ and $d$, which can simulate essentially any realistic disorder distribution in an approximate manner-- we note that $d=0$ implies that the charged impurities are simply located on the surface of the 3DTI.

There is considerable evidence \cite{Dohun_arXiv11} that in 3DTIs random unintentional charged impurities are the dominant source of disorder scattering.
We therefore assume that the quenched disordered potential $V_D$ is due to charged impurities.
For charged impurities, taking into account screening by the 2D surface carriers themselves in addition to the screening by the background lattice, we have
$n_{dis} |\langle V_{p{\bf k},p{\bf k'}}\rangle|^2 = n_{imp} |v_i(q)/\varepsilon(q,T)|^2$ where
$v_i (q) =2 \pi e^2/(\kappa q)$ is the Fourier transform of the 2D Coulomb
potential in a medium with an effective background static lattice dielectric constant $\kappa$, and
$\varepsilon(q,T)$ is the 2D static RPA dielectric function at finite temperature\cite{HwangDas_PRB07}.
The reported values of $\kappa$ for Bi$_2$Se$_3$ range from 30,
Ref.~\onlinecite{Beidenkopf_NPH11}, to $\approx 55$, Ref.~\onlinecite{Butch_PRB10}.

In Bi$_2$Se$_3$ the lowest optical phonon energy has been measured to
be $8.94$~meV$\approx 100$~K \cite{Kumar_PRB11,Richter_PSS77}
and therefore the optical phonons provide a substantial source of scattering
only at high temperature ($\gtrsim 250~K$) \cite{kim2012}.
Because we are interested primarily  only in the transport properties at temperatures
below 250~K in the remainder we neglect the contribution to the resistivity
due to optical phonons and consider only the contribution due to acoustic phonons.

Following references \onlinecite{Hwang_phononPRB08,Giraud_PRB11,Giraud_PRB12} we have that,
considering only longitudinal acoustic phonons,
the scattering time $\tau_{ph}$ is given by
\begin{equation}
\frac{1}{\tau_{ph}(\varepsilon)} = \sum_{\vk'}(1-\cos\theta_{\vk \vk'}) W_{\vk
  \vk'}\frac{1 - f(\varepsilon')}{1-f(\varepsilon)},
  \label{eq:phononscatter}
\end{equation}
where
\begin{eqnarray}
W_{\vk \vk'} = \frac{2\pi}{\hbar}\sum_{\vq}|C(\vq)|^2 &\times&
  [N_q \delta(\varepsilon-\varepsilon'+\omega_{\vq}) \nonumber \\
& + & (N_q + 1)
\delta(\varepsilon-\varepsilon'-\omega_{\vq})],
\label{wq}
\end{eqnarray}
is the transition probability from the state with momentum $\kk$ to the
state with momentum $\kk'$.
In \ceq{wq} $\vq = \vk-\vk'$,
$C(\vq)$ is the matrix element for scattering by acoustic phonon,
$\omega_{\vq}=v_{l} \vq$ is the acoustic phonon frequency with
$v_{l}$ the phonon velocity,
and $N_q = {1}/({\exp(\beta \omega_{\vq}) -1})$
is the phonon occupation number.
%The first (second) term ns Eq.~(\ref{wq}) corresponds to the
%absorption (emission) of an acoustic phonon of wave vector $\vq = \vk-\vk'$.
The matrix element $C(\vq)$ for the deformation potential electron-phonon coupling is given by
\begin{equation}
 |C(\vq)|^2 = \frac{D^2\hbar q}{2A\rho_m v_{l}}\left [ 1- \left (\frac{q}{2k} \right )^2 \right ]
 \label{eq:cq}
\end{equation}
where $D$ is the deformation potential coupling constant,
$A$ is the area of the sample,
and $\rho_m$
is the 2D mass density of one quintuple layer (around $1$ nm thick) of Bi$_2$Se$_3$,
given that the length scale over which the 2D surface states decay into the bulk
is approximately 1~nm \cite{LiuQiZh_PRB10}.

Using Eqs.~\ceq{eq:lcond}-\ceq{eq:cq} we can calculate the conductivity taking into
account scattering events due to both quenched disorder and phonons as long as the
system is homogeneous. However, especially close to the Dirac point, the random charged impurity induced disorder
potential causes the carrier density landscape to become strongly inhomogeneous, a fact
that has been observed experimentally in TIs \cite{Beidenkopf_NPH11} and previously
in graphene \cite{Martin_NP08,YZhang_NP09}.
To develop a theory in the presence of strong inhomogeneities it is first necessary
to characterize them. To do this we use the Thomas-Fermi-Dirac-Theory (TFDT)
first introduced in Ref.~\onlinecite{rossi2008}.
% that some of us have developed to calculate the disorder-averaged
%properties of the carrier density profile in the presence of a disorder potential.
In the TFDT, similarly to the Density Functional Theory (DFT), the energy of the
system is given by a functional of the density profile $n(\rr)$.
The great advantage
of a functional formalism is that it is not perturbative with respect to the spatial
fluctuations of the carrier density and therefore
can take into account nonlinear screening effects
that dominate close to the Dirac point. TFDT is just well-suited to describe the situation with large disorder-induced  spatial density inhomogeneity as in the low carrier density case whereas in the high-density situation, it simply gives the homogeneous density result with small fluctuations around the average density. In the TFDT, contrary to DFT, also the kinetic energy term is replaced
by a density functional.
This simplification makes the TFDT very efficient
computationally and therefore able to obtain disorder averaged quantities,
a task that cannot be accomplished using DFT. The simplification also makes
the TFDT in general less accurate than DFT \cite{polini2008}, however as long
as the characteristic length-scale over which the
density varies is larger than the local Fermi wavelength
$\lambda_F$, i.e. $|\nabla n/n|^{-1}\gg\lambda_F$, \cite{rossi2008,rossi2009,brey2009}
the TFDT returns reliable results \cite{rossi2008,dassarma2009,rossi2011}. Our results show that as in graphene \cite{rossi2008} the
condition $|\nabla n/n|^{-1}\gg\lambda_F$ is satisfied for the surface
of 3DTIs in typical experimental conditions. Close to the charge neutrality point (CNP)
 the density inside the electron-hole puddles is always different from zero (so that 
$\lambda_F$ is always finite) and of the order of $n_{rms}$. Our results show that $n_{rms} \sim n_{imp}$ and therefore the TFDT is also valid at the CNP as long as $n_{imp}$ is not too small. The great advantage of TFDT over DFT (to which TFDT is an approximation, as it uses the non-interacting kinetic energy functional) is that its relative numerical and computational ease enables one to use it for the calculation of transport properties using the computed ground state inhomogeneous spatial density profile, which would be completely computationally impossible for DFT to do.

Using the TFDT we can characterize completely the carrier density profile
in the presence of a disorder potential. We can obtain the typical length scale $L_D$ and root mean square fluctuation $\nrms$
of the disorder-induced carrier density inhomogeneities.
Using the Boltzmann theory
we obtain the relation between the mean free path and the doping, $\ell(n)= v_F\tau_{tot}(n)$, valid in the
homogeneous limit. In the limit in which $\ell(\nrms)\ll L_D$ the number of scattering events
inside a single homogeneous region, puddle, of the inhomogeneous landscape is large enough
that the Boltzmann theory is valid locally.
In addition, due to the
Klein tunneling, as in graphene
\cite{fogler2008,beenakker2008,katsnelson2006b,shytov2008,stander2009,young2009,rossi2010},
the resistance due to the boundaries between the puddles can be neglected in comparison
to the resistance arising from scattering events inside the puddles \cite{rossi2009}.
Under these conditions, due to the random distribution of the puddles, 2D transport
on the surface of a 3DTI can be described by the effective medium theory (EMT)
\cite{bruggeman1935,landauer1952,hori1975,rossi2009,fogler2009,dassarma2011}.
In the EMT, which is extensively used in science and engineering to quantitatively describe properties of highly inhomogeneous systems,  the conductivity of the inhomogeneous system is obtained as the
conductivity $\sigma_{EMT}$ of an equivalent homogeneous effective medium by averaging over disorder realizations
the local values $\sigma(n(\rr))$ given by the Boltzmann theory. The resulting implicit equation for $\sigma_{EMT}$ is:
\begin{equation}
 \int dn \frac{\sigma(n)-\sigma_{EMT}}{\sigma(n)+\sigma_{EMT}}P[n]=0
 \label{eq:emt}
\end{equation}
where $P[n]$ is the disorder-averaged carrier density probability distribution that we obtain
using the TFDT. A solution of the implicit EMT integral equation defined by Eq.~\ref{eq:emt} provides the effective conductivity of the inhomogeneous system.
In graphene the TFDT+EMT method has been shown to give results in remarkable agreement
with experiments \cite{dassarma2011} even in the highly inhomogeneous situation very close to the Dirac point and with full quantum transport analysis
\cite{rossi2012,rossi2010}.

A simplified approach that allows to  make further analytical progress and obtain
results in qualitative agreement with the numerical TFDT-EMT approach is the ``2-fluid''
model \cite{qzli_MLGPRB11}.
In this approach the inhomogeneous state characterized by the presence of electron-hole puddles
is approximated as a system comprised of
the ``electron-fluid'', formed by the electrons, and the ``hole-fluid'' formed by the holes,
with conductivity $\sigma_e$ and $\sigma_h$ respectively. Let $p$ be the fraction
of the system occupied by the electron gas, and consequently $(1-p)$ the fraction
occupied by the hole gas. Adapting Eq.~\ceq{eq:emt} to the simple case of only two
components for the inhomogeneous system we obtain the effective conductivity
\cite{kirkpatrick1973,Hwang_InsuPRB2010,qzli_MLGPRB11}:
\begin{equation}
\sigma_t = (p-\frac{1}{2})\left [ (\sigma_e -\sigma_h) +
  \sqrt{(\sigma_e-\sigma_h)^2+\frac{4\sigma_e \sigma_h}{(2p-1)^2}}
  \right].
\label{eq:sig_tot}
\end{equation}

One advantage of the 2-fluid model is that it allows to easily take into
account the effect of activation processes that at finite temperature,
especially close to the Dirac point, give a substantial contribution
to the conductivity and in particular qualitatively modify its temperature dependence. Carrier activation becomes operational when local potential fluctuations due to weak screening at low density lead to carrier confinement or localization in puddles, and global transport involves thermal activation of carriers over the local potential hills and barriers.  The activation process, which obviously becomes more important as the inhomogeneity becomes more important at lower carrier density, cannot be captured by the simple Boltzmann theory of Eq.~(\ref{eq:lcond}) or, for that matter, by any ensemble-averaged transport theory.  
To take into account the presence of activation processes $\sigma_e$ and $\sigma_h$
can be written as a sum of two terms
\cite{qzli_MLGPRB11}:
\begin{align}
 \sigma_e&=\sigmaena +\sigmaha \nonumber \\
 \sigma_h&=\sigmahna +\sigmaea \nonumber
 %\label{eq:sigma_na_a}
\end{align}
where $\sigmaena$, $\sigmahna$ are the disorder averaged conductivities obtained using the Boltzmann theory and
$\sigmaea$, $\sigmaha$ the contribution to the conductivity from activation processes.

$\sigmaena$, ($\sigmahna$) can be obtained by multiplying Eq.~\ceq{eq:lcond} by the ratio $n_e/n_0$ ($n_h/n_0$)\cite{qzli_MLGPRB11}.  Here, $n_e = \int_{-\infty}^{\infty} D_e (\epsilon) f(\eps)d\epsilon$ ($n_h = \int_{-\infty}^{\infty} D_h(\epsilon) [1-f(\epsilon)]d\epsilon$) denotes the effective electron (hole) density of inhomogeneous systems, while $n_0 = \int_{0}^{\infty} \frac{g_s g_v\epsilon }{2\pi(\hbar v_F)^2}f(\eps)d\epsilon$ is the electron density of homogeneous systems. The density of states $D_e(\epsilon)$ ($D_h(\epsilon)$) after disorder averaging are given by\cite{qzli_MLGPRB11}:
\begin{align}
 D_e(\epsilon) &= \int_{-\infty}^{\epsilon}\frac{g_s g_v(\epsilon - V)}{2\pi(\hbar v_F)^2}P(V) dV\nonumber\\
 D_h(\epsilon) &= \int_{\epsilon}^{\infty}\frac{g_s g_v(V - \epsilon)}{2\pi(\hbar v_F)^2}P(V) dV  \nonumber
\end{align}
% = \int_{-\infty}^{\infty} D_e f(\epsilon)d\epsilon
where $P(V)$ is the probability distribution of the screened disorder potential. We want to mention that we use the density of states for homogeneous systems in Eq.~\ceq{eq:lcond}, i.e., $D(\epsilon) = \frac{g_s g_v\epsilon }{2\pi(\hbar v_F)^2}$, to avoid double counting since the density inhomogeneity effects have already been considered through the variation of effective carrier density. At higher doping, both $n_e/n_0$ and the fraction of area occupied by electrons, denoting as $p = \int_{-\infty}^{E_F} P(V) dV$, approach unity while $n_h/n_0$ approaches zero. Note, however, that the ratio $n_e/n_0$ ($n_h/n_0$) being temperature-dependent is generally not equal to $p$ ($(1-p)$).
%For $P(V)$ we can use the TDFT results. However,
Given that the 2-fluid model is
an effective model the use of the exact $P(V)$ does not guarantee an increase
of its accuracy. It is more sensible to simply assume $P(V)$ to have an effective Gaussian
profile
\begin{equation}
 P(V) = \frac{1}{\sqrt{2\pi s^2}} \exp(-V^2/2s^2).
 \label{eq:fluc}
\end{equation}
with effective variance $s^2$.
The TFDT results, presented in  Sec.~\ref{sec:results}, show that the
root mean square $s$ of the screened disorder potential depends weakly on the doping or carrier density.
In Eq.~\ref{eq:fluc} we can then neglect the dependence of $s$ on the average
doping and use the value obtained for the Dirac point.
We consider two ways to estimate the value of the effective $s$ at the Dirac point that enters the Gaussian
approximation for $P(V)$:
({\bf i})  the {\em self-consistent} approximation introduced in Ref.~\onlinecite{adam2007};
({\bf ii}) the {\em quasi-TFDT}   approximation in which $s$ is fixed using
           the relation between $\nrms$ and $s$ in the Thomas-Fermi approximation
           with the value of $\nrms$ obtained from the full TFDT calculation.

In general $P(V)$ cannot be obtained analytically but it is possible to
obtain explicit expression for its moments \cite{galitski2007}.
For the second moment $\langle(\delta V)^2\rangle$
taking into account screening effects, within the random-phase-approximation (RPA)
for the surface states of  a 3DTI with total degeneracy $g=g_sg_v=1$
we have \cite{galitski2007}:
\beq
 \langle(\delta V)^2\rangle = 2 \pi n_{imp} \left(\frac{e^2}{\kappa}\right)^2 C_0 (r_s, a= k_F d)
 \label{eq:sniEn}
\eeq
where
%$d$ is the average distance of charged impurity to the 3D TI surface and the function $C_0$ is given by:
%
\begin{eqnarray}
 C_0 (r_s, a) &=& -1+\frac{e^{-4 a} r_s}{2+r_s}+\frac{64 E_1[4a]}{(8+\pi  r_s)^2}
 \\
 &+& e^{2 r_s a} (1+2 r_s a) (E_1 [2 r_s a]-E_1 [4a +2 r_s a]),\nonumber
 \end{eqnarray}
$r_s\equiv e^2/(\hbar v_F\kappa)$, and
$E_1 [z] = \int_z^{\infty} t^{-1}e^{-t}dt$ is the exponential integral function.
Assuming $P(V)$ to be a Gaussian with variance $s^2$ we have $s^2=\langle(\delta V)^2\rangle$.
The difficulty arises from the fact that the function $C_0$ depends on the density via the
Fermi wavevector $k_F$ that, in the presence of disorder, at the Dirac point cannot, because of fluctuations,
be taken to be simply zero. In the self-consistent approximation one
assumes that at the Dirac point the system can be approximated by a homogeneous
system having an effective carrier density $\hat n=(g/4\pi) \hat k_F^2$ such that
$\hat E_F^2 =s^2$, i.e. $\hbar^2v_F^2k_F^2=s^2$. Using this relation and
Eq.~\ceq{eq:sniEn} we obtain the following self-consistent equation for $s$:
\beq
 s^2 = 2 \pi n_{imp} \left(\frac{e^2}{\kappa}\right)^2 C_0 (r_s, a= s d/(\hbar v_F)).
 \label{eq:sCNPEn}
\eeq
By solving Eq.~\ceq{eq:sCNPEn} we obtain the value of $s$ within the self-consistent approximation.
In the {\em self-consistent+2-fluid model} the variance of the effective Gaussian
probability distribution $P(V)$ is obtained using Eq.~\ceq{eq:sCNPEn}.

By minimizing the Thomas-Fermi energy functional, at the Dirac point, for $g=1$, we obtain the
following equation
\beq
 2 \hbar v_F \sgn(n) \left(\frac{\pi  |n(\rr)|}{g_s g_v}\right)^{1/2} - V(\rr) = 0
 \label{eq:tf}
\eeq
where the first term is due to the kinetic energy and $V(\rr)$ is the local value of the screened disorder potential.
By disorder averaging \ceq{eq:tf} and assuming $P(V)$ to be a Gaussian with variance $s^2$ we obtain
\beq
 \nrms = \frac{g_s g_v}{2\pi (\hbar v_F)^2}\frac{\sqrt{3}}{4}s^2.
 \label{eq:nrms_vs_s}
\eeq
%
%xxxxxx
%xxxxxx
In the quasi-TFDT approximation the value of $s$ used in the 2-fluid model is set using
Eq.~\ceq{eq:nrms_vs_s} and the value of $\nrms$ obtained from the TFDT at the Dirac point.

Locally, the activated conductivities $\sigmaea$, $\sigmaha$ are given by:
\begin{align}
 \sigmaea(V(\rr)) &= \sigmaena\exp[\beta(E_F - V(\rr))] \\
 \sigmaha(V(\rr)) &= \sigmahna\exp[\beta(V(\rr) - E_F)].
\end{align}
where $V(\rr)$ is the local value of the screened disorder potential.
By disorder averaging these expressions and summing the contribution of the non-activated
conductivities we finally find \cite{qzli_MLGPRB11}:
\begin{eqnarray}
\sigma_e & = & \frac{1}{p}\int^{E_F}_{-\infty}(\sigmaena +
\sigma_h^{(a)})P(V) dV, \nonumber \\
&=& \sigmaena + \frac{\sigmahna}{2p} e^{
    \frac{\beta^2s^2}{2} -\beta E_F } {\rm erfc} \left (
    -\frac{E_F}{\sqrt{2} s} + \frac{\beta s}{\sqrt{2}} \right).
\label{eq:sige}
\end{eqnarray}
%Meanwhile the holes occupy the area with a fraction $q=1-p$ and
%the total conductivity of region 2 becomes
\begin{eqnarray}
\sigma_h & = &\frac{1}{(1-p)}\int_{E_F}^{\infty}(\sigmahna + \sigmaea)P(V) dV
\nonumber \\
&=& \sigmahna+\frac{\sigmaena}{2(1-p)} e^{
    \frac{\beta^2s^2}{2} +\beta E_F } {\rm erfc} \left (
    \frac{E_F}{\sqrt{2} s} + \frac{\beta s}{\sqrt{2}} \right).
\label{eq:sigh}
\end{eqnarray}
where the fraction of the system occupied by electrons is given by $p= \int_{-\infty}^{E_F} P(V) dV$  with Fermi energy $E_F$. These are the expressions of the electron and hole conductivities that enter the
2-fluid model whose total conductivity $\sigma_t$ is then obtained by using
Eq.~\ceq{eq:sig_tot}. We notice that in the expressions \ceq{eq:sige},~\ceq{eq:sigh} several
temperature effects are taken into account:
({\bf i})    the effect due to electron-phonon scattering processes that affect
             the total scattering time and therefore the value of $\sigmaena$ and $\sigmahna$;
({\bf ii})   the temperature dependence of the dielectric functions that
             enters in the calculation of $\tau_{dis}$ and that affect the value of $\sigmaena$ and $\sigmahna$;
({\bf iii})  the thermal broadening of the Fermi surface;
({\bf iv})   the presence of thermal activation.
The ability to capture all these effects makes the 2-fluid model a very useful
tool to study the transport properties of disordered TIs.

It is useful to explicitly write down the expression for the conductivity $\sigma_{CNP}$
at the charged neutrality point obtained using the 2-fluid model.
At the CNP we have that 50\% of the sample will be covered by electron puddles and
50\% by hole puddles so that $p=1-p=0.5$. In this case $\sigma_t=\sigma_e=\sigma_h$.
%Assuming $P(V)$ to be Gaussian
At zero temperature, there are no thermal activation effects and we find \cite{qzli_MLGPRB11}:
\beq
 \sigma_{CNP}=\sigma^{(na)}_{CNP}= \frac{1}{8G[r_s/2]}\frac{e^2}{h} \frac{g_s g_v}{2\pi(\hbar v_F)^2}\frac{s^2}{4 n_{imp}}
 \label{eq:sigma_na_cnp}
\eeq
where
\beq
\frac{G[x]}{x^2} = \frac{\pi}{4}+ 3x - \frac{3\pi x^2}{2}+\frac{x(3x^2 -2)\arccos[1/x]}{\sqrt{x^2-1}}
\eeq
Using Eq.~\ceq{eq:nrms_vs_s} we can rewrite \ceq{eq:sigma_na_cnp} in terms of $\nrms$:
\beq
 \sigma^{(na)}_{CNP} = \frac{1}{8G[r_s/2]}\frac{e^2}{h} \frac{\nrms}{\sqrt{3} n_{imp}}
 \label{eq:sigma_na_cnp2}
\eeq

In the following sections we present our results and in particular, when possible, a comparison
between the three approaches introduced: TFDT+EMT approach, quasi-TFDT+2-fluid model, and
self-consistent+2-fluid model.

%----------------

\section{Results}
\label{sec:results}
\subsection{TFDT results}
\label{sec:results:tfdt}
In this section we present the results for the carrier density distribution.
These results will then be used to calculate the
conductivity within the EMT and the 2-fluid models.
%shown in the following
%section.

\begin{figure}[htb]
\begin{center}
\includegraphics[width=0.99\linewidth]{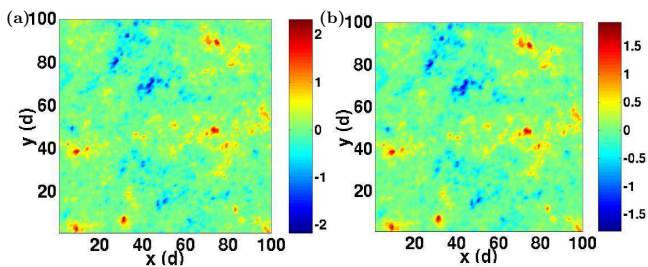}
  \caption{(Color online).  Color plots of carrier density distribution $n(\br)$ at the Dirac point for $n_{imp} = 7.5 \times 10^{13}$ cm$^{-2}$ and $\kappa = 50$. The color-scale is in units of $10^{13}$  cm$^{-2}$. (a) The impurity distance $d=0.1$ nm. (b) The impurity distance $d=0.2$ nm.}
\label{fig:En1}
\end{center}
\end{figure}

In Fig. \ref{fig:En1}, we show the carrier density profile calculated using the  TFDT
for a single disorder realization
with charged impurity density $n_{imp} = 7.5 \times 10^{13}$ cm$^{-2}$ and two different values
for the impurity distances, $d=0.1$~nm, and $d=0.2$~nm in panel (a) and (b) respectively.
Unless otherwise specified, we use the background dielectric constant $\kappa = 50$ for Bi$_2$Se$_3$
and the Fermi velocity $v_F = 6.4 \times 10^{5}$ m/s \cite{zhang2009ti,Dohun_arXiv11}. 
% The carrier density fluctuation is smaller for larger impurity distance $d$ to
%the surface of 3D TI because the Coulomb potential induced by the charged impurity
%is an exponential decreasing function of $d$.
%

\begin{figure}[htb]
\begin{center}
\includegraphics[width=0.9\linewidth]{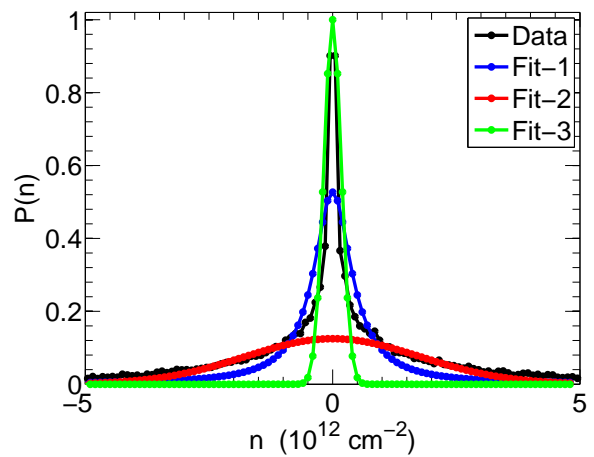}
  \caption{(Color online).  Density probability distribution at the Dirac point for $n_{imp} = 7.5 \times 10^{13}$ cm$^{-2}$, $d = 0.2$ nm and $\kappa = 50$.  The black line shows the TFDT result. The blue line is the fit $P = be^{-|n/a|}$ with $a = 0.7 \times 10^{12}$ cm$^{-2}$ and $b = 0.5309$. The red line is the Gaussian fit to the tails of $P (n)$, $P = be^{-x^2 /a^2}$ with $a = 2.5 \times 10^{12}$ cm$^{-2}$ and $b = .125$. The green line is the Gaussian fit to the center of $P (n)$, $P = be^{-x^2 /a^2}$ with $a = 0.25 \times 10^{12}$ cm$^{-2}$ and $b = 1.0.$}
  \label{fig:Pn}
\end{center}
\end{figure}

Figure \ref{fig:En1} conveys the nature of the carrier density landscape
on the surface of disordered 3DTIs as also shown recently by direct imaging
experiments \cite{Beidenkopf_NPH11}. To be able to make a quantitative
comparison with the experiments, and calculate the conductivity for large samples,
using the TFDT we calculate the disordered averaged density probability
distribution $P(n)$ for different parameter values like doping, $d$ and $\kappa$.
As in graphene \cite{rossi2008} we find that
$P(n)$ obtained using the TFDT is
bimodal, especially for finite values of the average carrier density, and so it
is not well fitted by any single curve.
%This is already clear from the
To exemplify this finding Fig. \ref{fig:Pn} shows $P(n)$ at the CNP
obtained using the TFDT and possible fitting curves. It is obvious
from the figure that a reasonable fit can be obtained only by using
two different Gaussian curves, one to fit the very high peak centered at
$n=0$ and one to fit the long tails of the distribution.
%
%at the CNP for $n_{imp} = 7.5 \times 10^{13}$ cm$^{-2}$ and $d = 0.2$ nm using the TFD theory. We use two different functions, {\it i.e.,} exponential and Gaussian functions, to fit the data. It is difficult to fit the $P[n]$ calculated with the TFD theory with one single Gaussian or exponential function and we need a bimodel distribution to capture the features of $P[n]$. The fact that the density distribution function $P[n]$ can not be fitted with a Gaussian approximation leads to the quantitative difference for the results of $\sigma_{min}$ between the TFD and Gaussian approximation approaches.

\begin{figure}[htb]
\begin{center}
\includegraphics[width=0.9\linewidth]{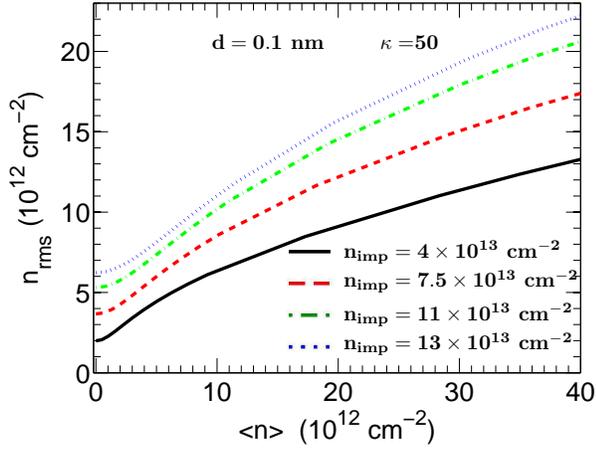}
  \caption{(Color online).  Root mean square of the density fluctuations $n_{rms}$ as a function of the average density for different values of $n_{imp}$ with $d=0.1$ nm and $\kappa = 50$. }
\label{fig:nrms1}
\end{center}
\end{figure}

The knowledge of $P(n)$ allows the calculation of all the statistical properties that characterize
the strongly inhomogeneous ground state of the surface of a 3DTI in the presence of disorder.
The quantity that better quantifies the strength of the carrier density inhomogeneities is
$\nrms$, shown in  Fig. \ref{fig:nrms1} as a function of the average carrier density $\langle n\rangle$ for different values of $\nimp$.
As expected larger values of $\nimp$ induce larger value of $\nrms$. The interesting result is that
$\nrms$ also increases with $\nav$ but the dimensionless ratio decreases with increasing average density. This is due to the fact that as $\nav$ increases the range of
values that $n$ can take locally also increases inducing larger values of $\nrms$, but the dimensionless ratio of the density fluctuation to the average density decreases with increasing density.

\begin{figure}[htb]
\begin{center}
\includegraphics[width=0.99\linewidth]{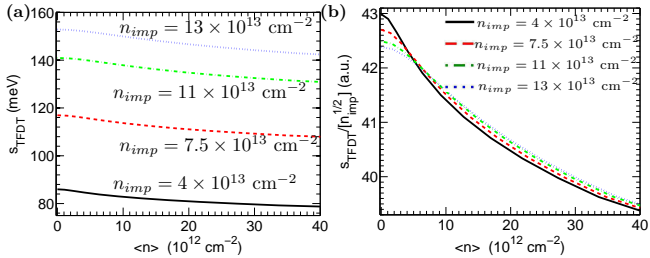}
  \caption{(Color online). (a) Root mean square of the screened disorder potential $s_{TFDT}$ as a function of average carrier density for different values of $n_{imp}$ with $d=0.1$ nm and $\kappa=50$ using TFDT. (b) $s_{TFDT}/\sqrt{n_{imp}}$ (in units of $\sqrt{10^{-13}}$meV$\cdot$cm) as a function of average carrier density.}
\label{fig:vsc1}
\end{center}
\end{figure}

Fig.~\ref{fig:vsc1}~(a) shows the TFDT results for the root mean square
of the screened disorder potential $s_{TFDT}$ as a function $\nav$
for different values of $\nimp$. As $\nav$ increases the screening
becomes more effective and therefore $s_{TFDT}$ decreases. However,
we see that the dependence of $s_{TFDT}$ on $\nav$ is fairly weak,
for all values of $\nimp$ considered. This fact justifies the
assumption in the quasi-TFDT model to assume the variance of the
effective distribution $P(V)$ to be independent of $\nav$.
Fig.~\ref{fig:vsc1}~(b) shows the scaling of the ratio
$s_{TFDT}/\sqrt{n_{imp}}$ versus $\nav$. We can see that
the ratio changes by less than 10\% over a wide range
of experimentally relevant values of the average density.
This result is one more evidence of the weak dependence
of $s_{TFDT}$ on $\nav$---to the zeroth order, the inhomogeneity and fluctuations are determined by the impurity distribution $n_{imp}$ and $d$ (as well as the background dielectric constant $\kappa$).
%
%From these results
%we can conclude that $s_{TFD}$ scales as $\sqrt{n_{imp}}$
%almost independent of $\nav$.
%

\begin{figure}[htb]
\begin{center}
\includegraphics[width=0.99\linewidth]{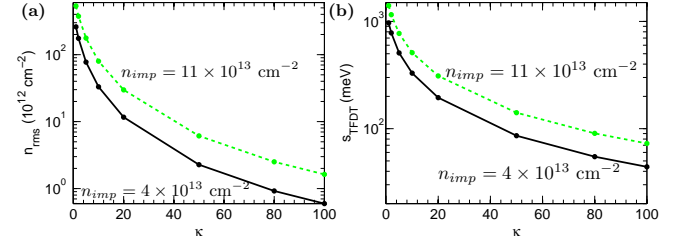}
  \caption{(Color online). TFDT results at the CNP for $d=0.1$ nm. (a) $n_{rms}$ as a function of substrate dielectric constant $\kappa$. (b) Root-mean-square of the screened disorder potential $s_{TFDT}$ as a function of substrate dielectric constant $\kappa$. The solid lines are for $n_{imp}= 4 \times 10^{13}$ cm$^{-2}$ and the dashed-lines are for $n_{imp}=11.0 \times 10^{13}$ cm$^{-2}$.}
\label{fig:vsc2}
\end{center}
\end{figure}

\begin{figure}[htb]
\begin{center}
\includegraphics[width=0.99\linewidth]{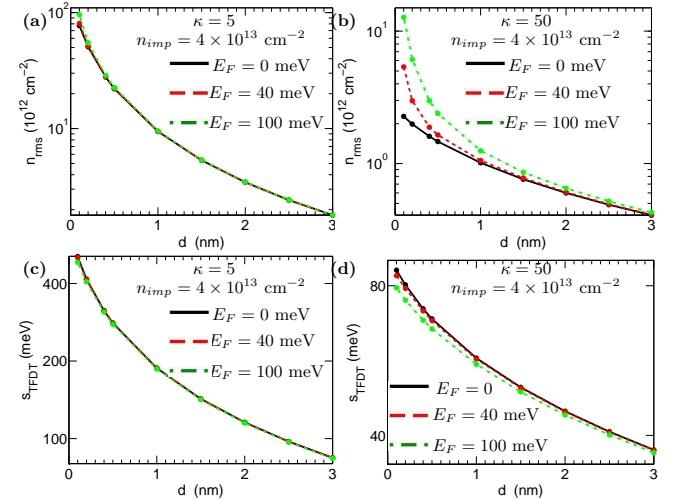}
  \caption{(Color online). (a) and (b) $n_{rms}$ at the Dirac point versus the distance $d$ of the charged impurity to the TI surface calculated within TFDT for $n_{imp}=4.0 \times 10^{13}$  cm$^{-2}$. The solid, dashed and dot-dashed lines correspond to chemical potential $E_F= 0$, $40$ and $100$ meV, respectively. (c) and (d) The corresponding $s_{TFDT}$ results versus $d$. (a) and (c) The effective background dielectric constant $\kappa=5$. (b) and (d) The effective background dielectric constant $\kappa=50$.}
\label{fig:sd}
\end{center}
\end{figure}

The dependence of $\nrms$ and $s_{TFDT}$ on $\kappa$ and  $d$ is very strong as shown
by Fig.~\ref{fig:vsc2} and Fig.~\ref{fig:sd}.
From Fig.~\ref{fig:vsc2}  we see that both $\nrms$ and $s_{TFDT}$ decrease rapidly
as $\kappa$ increases. Increasing the average distance of the charged impurities
from the surface of the 3DTI also strongly reduces the amplitude of the spatial
inhomogeneities and therefore of $\nrms$ and $s_{TFDT}$ as shown in Fig.~\ref{fig:sd}.

\begin{figure}[htb]
\begin{center}
\includegraphics[width=0.99\linewidth]{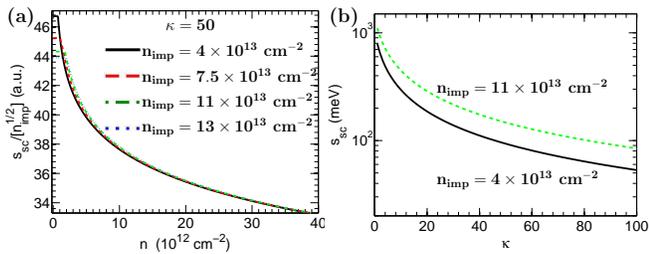}
  \caption{(Color online). Self-consistent results. (a) $s_{sc}/\sqrt{n_{imp}}$ (in units of $\sqrt{10^{-13}}$meV$\cdot$cm) as a function of carrier density for different values of $n_{imp}$ with $\kappa = 50$ and $d=0.1$ nm. (b) Potential fluctuation $s_{sc}$ as a function of substrate dielectric constant $\kappa$. The solid lines are for $n_{imp}= 4 \times 10^{13}$ cm$^{-2}$ and the dashed-lines are for $n_{imp}=11.0 \times 10^{13}$ cm$^{-2}$ with $d=0.1$ nm.}
\label{fig:scas}
\end{center}
\end{figure}

We now present a comparison of the TFDT results with the ones obtained using
the self-consistent and the quasi-TFDT approaches, methods in which
$P(V)$ is assumed to be Gaussian.
Fig.~\ref{fig:scas} shows the scaling of the screened disorder rms
obtained using the self-consistent approach ($s_{sc}$) versus doping and $\kappa$.
These results, analogously to the TFDT results,
show that $s$ depends very weakly on $\nav$ and quite strongly on $\kappa$.

\begin{figure}[htb]
\begin{center}
\includegraphics[width=0.99\linewidth]{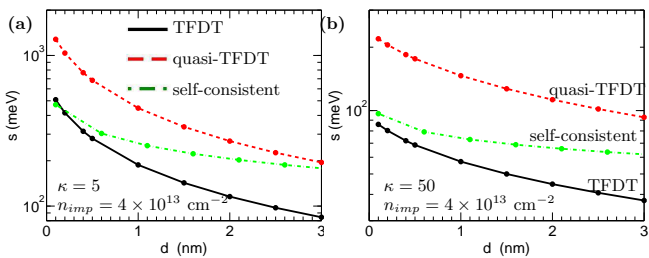}
  \caption{(Color online). Comparison of three methods for the potential fluctuation $s$ at the Dirac point versus the charged impurity distance $d$ to the TI surface for $n_{imp}=4.0 \times 10^{13}$  cm$^{-2}$. The solid, dashed and dot-dashed lines correspond to the TFDT, quasi-TFDT and the self-consistent methods, respectively.  (a) The effective background dielectric constant $\kappa=5$. (b) The effective background dielectric constant $\kappa=50$.}
\label{fig:sd2}
\end{center}
\end{figure}

Fig.~\ref{fig:sd2} shows the comparison for the value of $s$ at the CNP as a function of $d$ obtained using the
three different methods: TFDT, quasi-TFDT and, self-consistent. We see that the self-consistent approach in general
returns values of $s$ larger than the ones obtained using the TFDT. We should emphasize that $s$
obtained using the quasi-TFDT method is only an effective quantity that is used to calculate the transport
properties within the 2-fluid model.
% and that is not surprising that is close to $s_{TFDT}$ given that
%to obtain it he value of $\nrms$ from the full TFDT is used.
%

%-----------------------------

\subsection{Transport at zero temperature}
\label{sec:zeroTcond}
\begin{figure}[htb]
\begin{center}
\includegraphics[width=0.9\linewidth]{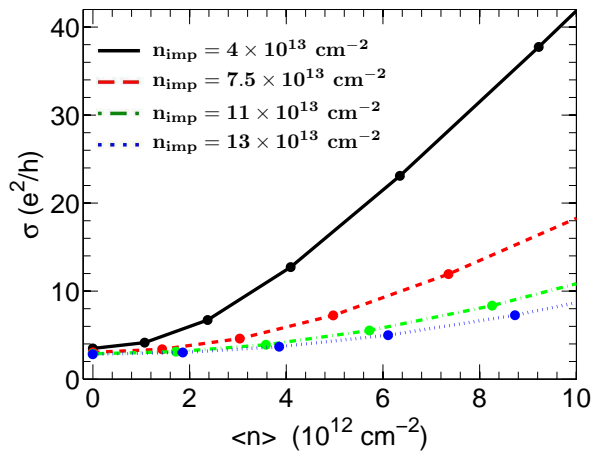}
  \caption{(Color online). TFDT-EMT conductivity as a function of average carrier density for different values of  $n_{imp}$ with $d=0.1$ nm and $\kappa = 50$. }
    \label{fig:sigmaN}
\end{center}
\end{figure}

Using the TFDT and the EMT we can calculate the 2D conductivity on the surface of a 3DTI.
Fig.~\ref{fig:sigmaN} shows $\sigma$ as a function of doping, i.e. average density, for several values of
$\nimp$ and fixed $d=0.1$~nm and $\kappa=50$.
As in graphene \cite{dassarma2011} the TFDT+EMT results recover the behavior
of $\sigma(n)$ observed experimentally \cite{Dohun_arXiv11}:
the linear scaling of $\sigma(n)$ at large doping, the finite value ($\sigma_{\rm min}$) of $\sigma$
for $n=0$, and the crossover regime for intermediate values of $n$.
More importantly, the theory returns values of $\sigma_{\rm min}=(2-4)e^2/h$
(depending on the sample properties) that agree with the ones observed
in experiments \cite{Dohun_arXiv11}.

\begin{figure}[htb]
\begin{center}
\includegraphics[width=0.99\linewidth]{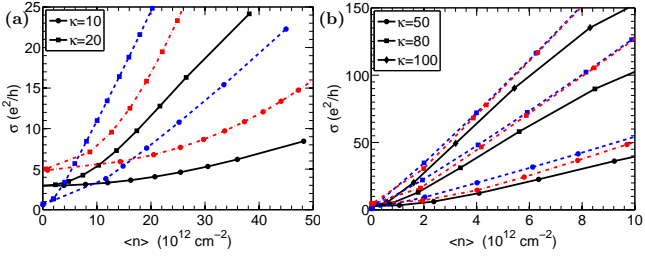}
  \caption{(Color online).  Calculated conductivity as a function of carrier density for different values of the substrate dielectric constant with the impurity density $n_{imp} = 4.0 \times 10^{13}$ cm$^{-2}$ , $d = 0.1$ nm.  The solid, dashed and dot-dashed lines are obtained by using TFDT-EMT method, self-consistent method, and quasi-TFDT method, respectively. }
  \label{fig:signni4}
\end{center}
\end{figure}

Fig.~\ref{fig:signni4} shows the comparison for the dependence of $\sigma$ with respect to $n$ obtained using the
TFDT+EMT method and the 2-fluid approximations.
%As described in Sec.~\ref{sec:theory} in the 2-fluid model we use an effective Gaussian probability distribution for the screened disorder potential, $P(V)$,
%with standard deviation $s$. The value of $s$ can be estimated using the self-consistent approach
%\cite{adam2007} or the relation between $s$ and $\nrms$ obtained from the Thomas-Fermi theory and
%$\nrms$ from the full TFDT (quasi-TFDT approach). Depending on the approximation used we have two 2-fluids models:
%the ``self-consistent+2-fluids'' model, and the ``quasi-TFDT+2-fluids'' model.
From Fig.~\ref{fig:signni4} we see that for $n_{\rm imp}\lesssim 4\times 10^{13}$~cm$^{-2}$ and values of $\kappa\gtrsim 50$,
the three approaches give very similar results. For smaller values of $\kappa$ the three approaches give results
that are qualitatively similar but that differ quantitatively. The general conclusion is that away from the CNP the three
approaches agree quantitatively for samples with mobility $\mu> 1000 {\rm cm}^2/{\rm V\cdot s}$. For very low mobility highly-disordered TI systems (with 2D surface mobility lower than $1000$ cm$^2$/V$\cdot$s even at $T=0$ ), we expect the TFDT to provide the most quantitatively accurate results. However,
given that our main objective is to describe the universal qualities of the transport arising from the presence of inhomogeneities, and given the absence of accurate knowledge of the impurity distribution, i.e. $n_{imp}$ and $d$ in our model, for the purposes of this work the three approaches appear to be equivalent.

\begin{figure}[htb]
\begin{center}
\includegraphics[width=0.9\linewidth]{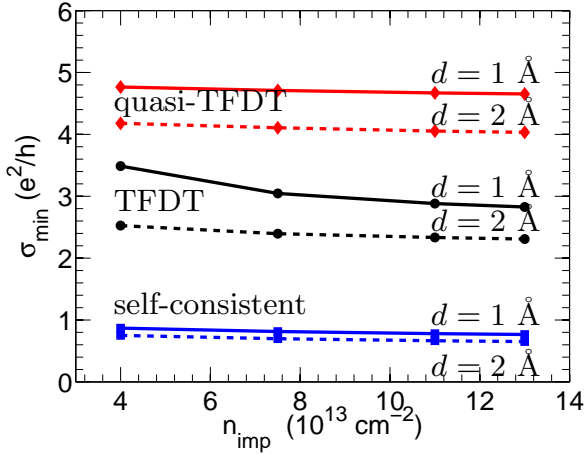}
  \caption{(Color online). The minimum conductivity $\sigma_{min}$ as a function of charged impurity density $n_{imp}$ with $\kappa = 50$. The symbols diamond, circle and square denote the results obtained using the quasi-TFDT, TFDT-EMT and the self-consistent method, respectively. The solid and dashed lines correspond to the impurity distance $d=1$ \AA \ and $d=2$ \AA \ , respectively. }
  \label{fig:sig_min_nimp}
\end{center}
\end{figure}

\begin{figure}[htb]
\begin{center}
\includegraphics[width=0.9\linewidth]{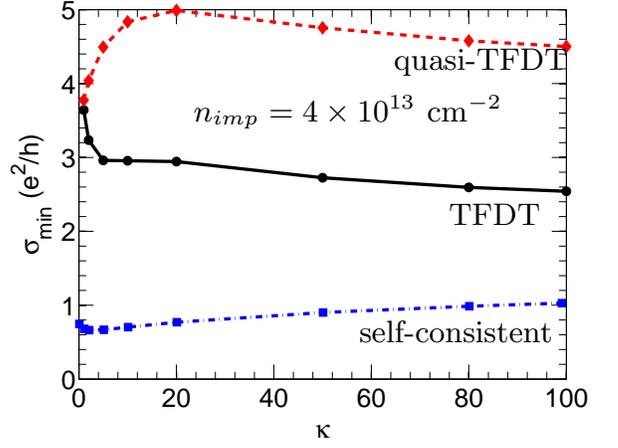}
  \caption{(Color online).   The minimum conductivity $\sigma_{min}$ as a function background dielectric constant $\kappa$ for $n_{imp} =4.0 \times 10^{13}$ cm$^{-2}$ , $d = 0.1$ nm. The dashed, solid and dot-dashed lines denote the results obtained using the quasi-TFDT, TFDT-EMT and the self-consistent method, respectively. }
  \label{fig:sig_ni4_eps}
\end{center}
\end{figure}

\begin{figure}[htb]
\begin{center}
\includegraphics[width=0.99\linewidth]{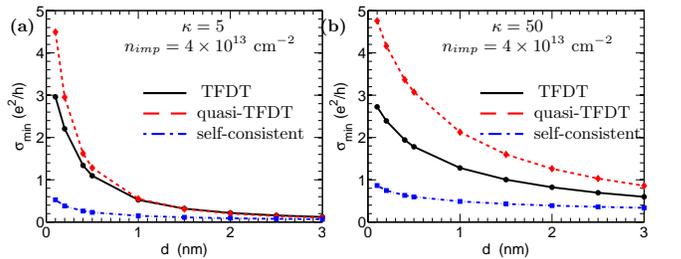}
  \caption{(Color online).   The minimum conductivity $\sigma_{min}$  versus the charged impurity distance $d$ to the TI surface for $n_{imp}=4.0 \times 10^{13}$  cm$^{-2}$. The solid, dashed and dot-dashed lines correspond to the TFDT-EMT, quasi-TFDT and the self-consistent methods, respectively. (a) The effective background dielectric constant $\kappa=5$. (b) The effective background dielectric constant $\kappa=50$.}
  \label{fig:sigmind}
\end{center}
\end{figure}

Close to the CNP the three transport approaches give results that differ quantitatively
as shown in Figs.~\ref{fig:sig_min_nimp}-\ref{fig:sigmind}. The self-consistent+2-fluid model
for the parameter values relevant for TIs gives values of $\smin$ smaller than the ones obtained
using the TFDT+EMT method. In TIs the agreement between the two methods close to the CNP
is worse than in graphene \cite{Adam_SCC09} due to the fact that in TIs the density of charged
impurities is larger than in typical graphene samples.
It is remarkable to see how, in agreement with experiments, the three methods give that $\smin$
depends very weakly on $\nimp$ and $\kappa$,
for experimentally relevant parameter values.
The weak dependence on $\nimp$ can be qualitatively understood using the 2-fluid model result for
$\smin$ Eq.~\ceq{eq:sigma_na_cnp}
from which we see that
$\smin$ is proportional to the ratio $s^2/n_{imp}$, and that on the other hand $s^2$
is proportional to $\nimp$, Eq.~\ceq{eq:sCNPEn}.
%and therefore $\smin$ depends very weakly on $$\nimp$$.
%Similarly In addition from Eqs.~\ceq{eq:sgmin},~\ceq{eq:selfcons} we see that dependence on $\kappa$,
%via $r_s$ and the functions $G(r_s/2)$ and $C_0(r_s,s d/(\hbar v_F))$ is also very weak.
%
Within the semiclassical approach the very weak dependence of $\smin$
on $\nimp$ and $\kappa$ is due to the fact that an increase (decrease) of $\nimp$ ($\kappa$)
increases the strength of the disorder potential that causes a decrease of the carriers mean
free path $\ell$, and an increase of the amplitude of the carrier density inhomogeneities(i.e. the
density of carriers in the electron-hole puddles). The reduction of $\ell$ and the increase
of $\nrms$ have opposite effect on $\smin$ and they almost cancel out
\cite{jang2008,rossi2009}.

The dependence of $\smin$ with respect to the average distance $d$ of the charged impurities from the TI's surface
is appreciably different for the three methods. The self-consistent+2-fluid model returns a very weak dependence
of $\smin$ with respect to $d$. This is due to the weak dependence with respect to $d$ of $s^2$ obtained
using the self-consistent approximation, Eq.~\ceq{eq:sCNPEn}.
On the other hand the value of $s^2$ obtained using the TFDT is quite sensitive to the value of $d$ and as a consequence
using the TFDT+EMT and the quasi-TFDT+2-fluid models we find that the dependence of $\smin$ on $d$ is not as weak
as the one given from the self-consistent+2-fluid model.
All the three approaches show that $\smin$ decreases as a function of $d$. This is due to the fact that as $d$ increases,
due to the weakening of the disorder potential, the decrease of $\nrms$ is faster than the increase of the mean free path.
%

%\clearpage
%-----------------------------

\subsection{Transport at finite temperature}
\label{sec:finiteTcond}
\begin{figure}[htb]
\begin{center}
\includegraphics[width=0.9\linewidth]{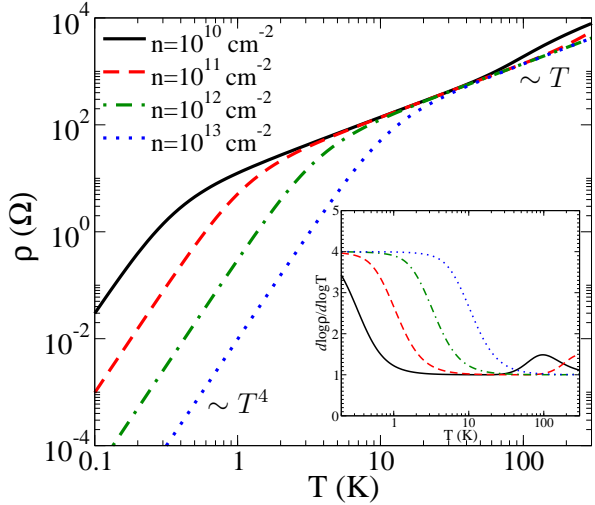}
  \caption{(Color online). Double-logarithmic scale of resistivity (Bi$_2$Se$_3$)  only due to phonon scattering as a function of temperature for several densities. The deformation potential coupling constant $D= 30$ eV, $\rho_m \simeq 7.68 \times 10^{-7}$ g/cm$^{-2}$, and the velocity of longitudinal acoustic phonon mode $v_l = 2900$ m/s. The inset shows the logarithmic derivatives $d \log \rho/d\log T$ versus temperature. }
\label{fig:OnlyPhonon}
\end{center}
\end{figure}

In this section, using the 2-fluid model, we present our results for the conductivity in TIs at finite
temperature.
%At finite temperatures the transport properties are affected by the electron-scattering
%processes and by activation processes.
% that in Dirac's materials play a very important role
%especially close to the CNP.
%
If we neglect the contribution of activation processes in TIs the dominant contribution
to the temperature dependence of $\sigma$ is due to electron-phonon scattering processes.
Fig.~\ref{fig:OnlyPhonon} shows the longitudinal acoustic phonon limited resistivity of Bi$_2$Se$_3$ surface as a
function of temperature on a log-log plot.
To calculate the Bi$_2$Se$_3$ surface resistivity due to phonon scattering\cite{Giraud_PRB12}
we use $D=30$~eV for the deformation potential coupling constant,
$\rho_m \simeq 7.68 \times 10^{-7}$ g/cm$^{-2}$ for
the two dimensional mass density, i.e. 1 quintuple layer mass density of Bi$_2$Se$_3$,
and $v_l = 2900$~m/s is the velocity of the longitudinal acoustic phonon mode\cite{Giraud_PRB12}.
The inset of Fig. \ref{fig:OnlyPhonon} shows the logarithmic derivatives of the temperature-dependent resistivity.
Fig. \ref{fig:OnlyPhonon} clearly demonstrates two different regimes depending on whether the phonon system is degenerate or non-degenerate, and the low- to high-temperature crossover is characterized by the Bloch-Gr\"uneisen (BG) temperature $T_{BG} = 2 k_F v_l/k_B$\cite{Hwang_phononPRB08,hongkihwang_PRB11,Giraud_PRB11,Giraud_PRB12}.
The resistivity increases with $T$ as $\rho \sim T^4$ at low temperatures and $\rho \sim T$ at high temperatures, which agrees with the results obtained for TI films by using an isotropic elastic continuum approach\cite{Giraud_PRB12}. We note that electron-phonon scattering is an important scattering mechanism
for finite-temperature transport in TIs, e.g., $T_{BG} \sim 30$ K in TIs compared to $T_{BG} \sim 100$ K in graphene
because the latter has much larger phonon velocity. In graphene, in contrast to 2D surface TI transport, phonon effects are extremely weak\cite{Hwang_phononPRB08}.

\begin{figure}[htb]
\begin{center}
\includegraphics[width=0.99\linewidth]{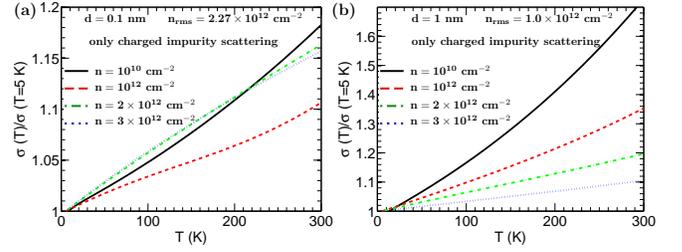}
\caption{(Color online). Calculated conductivity  as a function of temperature for different carrier density including only charged impurity scattering with $d=0.1$ nm, $\kappa=50$ and $n_{imp}= 4  \times 10^{13}$ cm$^{-2}$. (a) The charged impurity distance $d=0.1$ nm. The corresponding TFDT $n_{rms} \simeq 2.27 \times 10^{12}$ cm$^{-2}$ and the potential fluctuation $s \simeq 220$ meV (Eq.~\ref{eq:nrms_vs_s}). (b) The charged impurity distance $d=1$ nm.  The corresponding TFDT $n_{rms} \simeq 1.0 \times 10^{12}$ cm$^{-2}$ and the potential fluctuation $s \simeq 150$ meV (Eq.~\ref{eq:nrms_vs_s}). }
  \label{fig:sigTn2}
\end{center}
\end{figure}

The results shown in Fig.~\ref{fig:OnlyPhonon} do not include the contribution to the resistivity
due to quenched disorder. To include both the effect of quenched disorder
and electron-phonon scattering we use the  quasi-TFDT+2-fluid.
As discussed in Sec.~\ref{sec:theory}, this approach
allows us to take into account several finite temperature effects:
the temperature dependence of the screening of the quenched disorder,
electron-phonon scattering processes, broadening of the Fermi surface,
and temperature induced activated processes.
The temperature activated processes cause
$\sigma$ to increase with $T$ and therefore induce an insulating behavior (i.e. conductivity increasing with increasing temperature)
for  $\sigma(T)$, whereas the electron-phonon scattering processes induce
a metallic behavior (i.e. conductivity decreasing with increasing temperature). The change with $T$ of the screening of the disorder
potential also induces a metallic behavior for $\sigma(T)$, however
given the large value of $\kappa$ in typical TIs this effect is quite weak,
contrary to the case of graphene
\cite{dassarma2011} or 2D semiconductor systems\cite{dassarma1999}. Figs.~\ref{fig:sigTn2}~(a),~(b) show the dependence of $\sigma$
on $T$ obtained by neglecting the effect of electron-phonon scattering
processes.
We see that in this case the temperature dependence of $\sigma$ is almost
completely determined by thermally activated processes that induce a
monotonically increase of $\sigma$ with $T$.

\begin{figure}[htb]
\begin{center}
\includegraphics[width=0.99\linewidth]{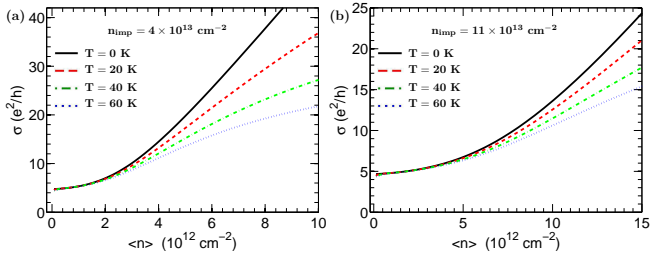}
\caption{(Color online). Calculated conductivity as a function of net carrier density for various temperatures with $d=0.1$ nm and $\kappa=50$. (a) The impurity  density  $n_{imp}=4  \times 10^{13}$ cm$^{-2}$. The corresponding TFDT $n_{rms} \simeq 2.27 \times 10^{12}$ cm$^{-2}$ and the potential fluctuation $s \simeq 220$ meV (Eq.~\ref{eq:nrms_vs_s}); (b) The impurity  density  $n_{imp}=11  \times 10^{13}$ cm$^{-2}$. The corresponding TFDT $n_{rms} \simeq 4.21 \times 10^{12}$ cm$^{-2}$ and the potential fluctuation $s \simeq 300$ meV (Eq.~\ref{eq:nrms_vs_s})}
  \label{fig:Ph_sigNT}
\end{center}
\end{figure}

\begin{figure}[htb]
\begin{center}
\includegraphics[width=0.99\linewidth]{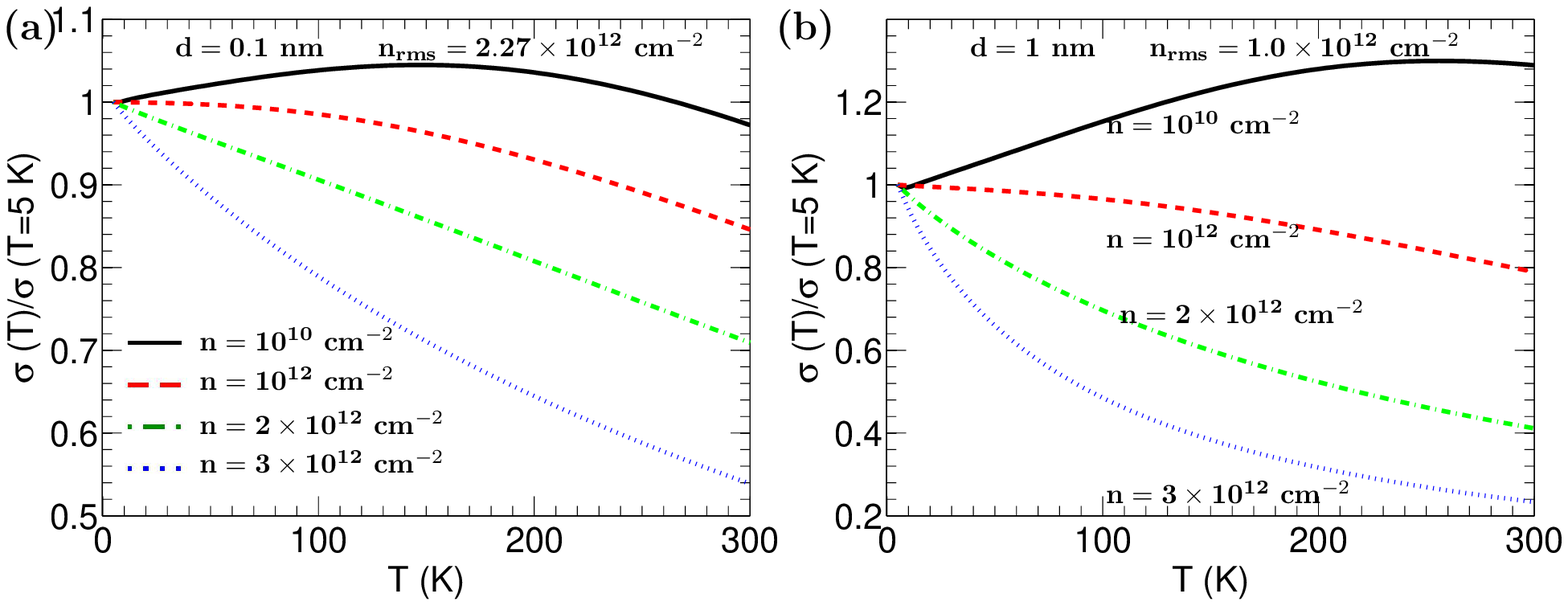}
\caption{(Color online). Calculated conductivity  as a function of temperature for different carrier density including both charged impurity and electron-phonon scattering, with $d=0.1$ nm, $\kappa=50$ and $n_{imp}= 4  \times 10^{13}$ cm$^{-2}$. (a)  The charged impurity distance $d=0.1$ nm. The corresponding TFDT $n_{rms} \simeq 2.27 \times 10^{12}$ cm$^{-2}$ and the potential fluctuation $s \simeq 220$ meV (Eq.~\ref{eq:nrms_vs_s}). (b) The charged impurity distance $d=1$ nm.  The corresponding TFDT $n_{rms} \simeq 1.0 \times 10^{12}$ cm$^{-2}$ and the potential fluctuation $s \simeq 150$ meV (Eq.~\ref{eq:nrms_vs_s}). }
  \label{fig:sigTn}
\end{center}
\end{figure}

\begin{figure}[htb]
\begin{center}
\includegraphics[width=0.99\linewidth]{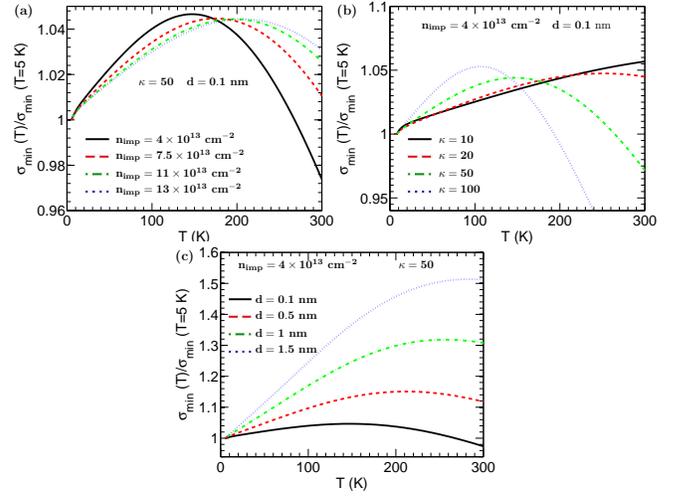}
\caption{(Color online). Minimum conductivity as a function of temperature with (a) for various charged impurity density $n_{imp}$. As $n_{imp}$ increases, the corresponding $s =  220, \ 300, \ 360, \ 390$ meV; (b) for various dielectric constant $\kappa$. As $\kappa$ increases, the corresponding $s = 840, \ 500, \ 220, \ 110$ meV; (c) for various charged impurity distance $d$. As $d$ increases, the corresponding $s = 220, \ 180, \ 150, \ 130$ meV. The corresponding potential fluctuation $s$ is obtained via Eq.~\ref{eq:nrms_vs_s} using $n_{rms}$ calculated within the TFDT (shown in Figs.~\ref{fig:nrms1}, \ref{fig:vsc2} and \ref{fig:sd}).}
  \label{fig:sigmintot}
\end{center}
\end{figure}

Fig.~\ref{fig:Ph_sigNT} shows
the scaling of the conductivity with respect to doping at different temperatures including
both the effects of quenched disorder and electron-phonon scattering.
At large densities the main effect of the finite temperature is to
suppress $\sigma$ due to the presence of electron-phonon scattering processes.
However, at low densities the effect of electron-phonon scattering processes competes
with thermal activation processes and can give rise to a non-monotonic
dependence of $\sigma$ with respect to $T$.
This is shown in Figs.~\ref{fig:sigTn}~(a),~(b) where $\sigma(T)$
for different values of $n$ is plotted.
%taking into account only the thermal activation processes.
From Figs.~\ref{fig:sigTn}~(a),~(b) we see that for $n\lesssim n_{\rm rms}$
at low temperatures the thermal activation processes
dominate and induce an insulating behavior for $\sigma(T)$.
The crossover temperature from insulating to metallic behavior
depends on $n$, $\nimp$, $\kappa$ and $d$.
In general the larger the strength of the spatial fluctuations
of the carrier density
the stronger is the effect of thermal activation
processes and therefore the larger is the low temperature range
for which the transport exhibits insulating behavior.
This is shown clearly by the scaling of $\sigma(T)$ at the CNP
for different values of $\nimp$, $\kappa$,
Figs.~\ref{fig:sigmintot}~(a), (b), respectively.
From Figs.~\ref{fig:sigmintot}~(a) and (b) we see that a change of $\nimp$ and $\kappa$
that increases $\nrms$ extends
the
range of temperatures over which $\sigma_{\rm min}(T)$ exhibit
an insulating behavior.

%Panels (a), (b) and (c) of these figure respectively show that
%a change of \nimp, $\kappa$ or $d$ that increase $\nrms$
%increases the
%range of temperatures over which $\sigma_{\rm min}(T)$ exhibit
%an insulating behavior.
% {\bf ???? Behavior with respect to d not completely understood}

The temperature dependence of $\sigma$ for different values of $d$ is shown in Fig.~\ref{fig:sigmintot}~(c). It appears to contradict the general rule that a parameter
change that increases $\nrms$ will increase the range
of temperatures over which $\sigma(T)$ exhibit an insulating
behavior. This is due to the combination of two effects:
({\bf i})  The fact that at large $d$ ($d\gtrsim 1$~nm) $\nrms$
           is very low and so
           the resistivity $\rho_{dis}$  due to the quenched
           disorder at low carrier densities is much higher than the resistivity
           $\rho_{ph}$ due to electron-phonon scattering,
           so that $\sigma_{tot}(T)/\sigma_{tot}(T=5 K)\approx\sigma_{dis}(T)/\sigma_{dis}(T=5 K)$;
({\bf ii}) The decrease at large $d$ of the metallic screening effects.
The scaling of $\sigma$ with respect to $T$ and $d$ is therefore very interesting
because it reveals the temperature dependence of the screening and could therefore
be used to indirectly identify the nature of the disorder potential and the
screening  properties of the surfaces of 3DTIs.


Before concluding this section on the temperature-dependent surface conductivity of 3DTIs, we point out that one of the important qualitative findings of our work is the nonmonotonic temperature dependence of the 2D surface conductivity as apparent in Figs.~\ref{fig:Ph_sigNT}-\ref{fig:sigmintot} and as expected from the competing mechanisms of phonon scattering and disorder induced density inhomogeneity.  In particular, phonons induce higher-temperature metallic temperature dependence and the density inhomogeneity induces insulating temperature dependence through thermal activation, and at some disorder-dependent (and also doping-dependent)  characteristic temperature the transport behavior changes from being insulating-like to metallic-like.  We emphasize that this temperature-induced crossover behavior has nothing to do with any localization phenomenon (and in fact,  Anderson localization effects are completely absent in TI surface transport since all back scattering is suppressed), and it arises entirely from a competition between inhomogeneity and phonons.  The metallic behavior moves to higher (lower) temperature as disorder increases (decreases).  This nonmonotonicity has been observed in experiments at lower 2D carrier densities\cite{kim2012} where the inhomogeneity effects are important.

%\clearpage

%-----------------------------

\section{Conclusions}
\label{sec:conclusions}
We have presented a detailed theoretical study of the 2D transport properties of the surfaces of 3DTIs.
There is compelling evidence that in current transport experiments on 3DTIs charged impurities
are the dominant source of quenched disorder. For this reason in our study we have considered
in detail the case in which the quenched disorder potential is the one created by random
charged impurities close to the surface of the TI.
However, the theoretical framework that we have developed and presented in this work is very general.
As in graphene, the presence of charged impurities induces the formation of strong carrier density
inhomogeneities. In particular, close to the charge neutrality point the carrier density landscape
breaks-up in electron-hole puddles. The strong carrier density inhomogeneities make the theoretical
description of the electronic transport challenging for two reasons:
({\bf i})  inability to use standard theoretical methods that assume a homogeneous density landscape;
({\bf ii}) the importance at finite temperature of thermally activated processes.
Our work presents a theoretical description of transport on the surface of 3DTIs
that overcomes these difficulties and takes into account also the effect of electron-phonon scattering processes which is an important resistive mechanism at higher temperatures.

To characterize the disorder-induced inhomogeneities we use the Thomas-Fermi-Dirac-Theory. We also
present a comparison of the results obtained using the TFDT with the one obtained using the
self-consistent approximation and the quasi-TFDT approach and show that the three approaches
give results that are qualitatively similar but that differ quantitatively.

To study the electronic transport in the presence of strong carrier density inhomogeneities,
starting from the TFDT results, we use the effective medium theory and a 2-fluid model.
The TFDT+EMT approach is well justified and is expected to provide the most accurate
results. However, the generalization of the TFDT+EMT method to finite temperature is impractical
due to the contribution to transport of thermally activated processes, contribution that
can be dominant at low temperatures due to the strong carrier density inhomogeneities.
At finite temperature the 2-fluid approach is very valuable because it allows to
take into account all the finite temperature effects, such as temperature dependent
screening and electron-phonon scattering, including thermally activated processes.
The 2-fluid model relies on the use of  an effective Gaussian distribution for the probability distribution of
the screened potential $P(V)$. The parameters that define $P(V)$ can be chosen
in such a way to maximize the agreement of the results for $T=0$ obtained using the 2-fluid model
and the ones obtained using the TFDT+EMT approach. These parameters are then
used to obtain the transport properties at finite temperature using the 2-fluid model.

In current 3DTIs the dielectric constant ($\kappa\sim 50$) is much larger than in graphene
where $\kappa\sim 1-4$. In addition, in 3DTIs the acoustic phonon velocity is smaller than in graphene.
These facts make the contribution of electron-phonon scattering processes to the resistivity
much more important in the surface of 3DTIs than in graphene. For the surfaces of 3DTIs
the effect of electron-phonon scattering events becomes important already for $T$ as low
as 10~K, whereas in graphene it becomes relevant only for $T\gtrsim 200$~K.
As a consequence for the surfaces of 3DTIs we find that electron-phonon scattering
is much more important to determine the dependence of the conductivity on $T$.
The large value of $\kappa$ in 3DTIs also implies that
for the surface of 3DTIs, contrary to graphene, the temperature dependence of the
screening does not play an important role.
The temperature dependence of the conductivity on the surface of 3DTIs is therefore mostly
determined by electron-phonon scattering processes and thermal activations processes.
These two types of processes have opposite effect on $\sigma$ and at low temperature
and low doping compete
giving rise to a nonmonotonic dependence of $\sigma$ with respect to $T$. The nonmonotonic temperature dependence of the 2D surface conductivity is one of the important new qualitative results of our theory.
We have presented detailed results for $\sigma(T)$ that clearly show
the competition of the different processes that affect $\sigma$ for $T\neq 0$.

The theoretical approach developed here, being able to include all the main effects
that determine the transport properties of the surfaces of 3DTIs, allowed us to
present results that can be directly and quantitatively compared to the experimental ones.
The good agreement between our theoretical results and the recent experimental
measurements
\cite{Dohun_arXiv11,Hong_arXiv11,Kong_Natnano,Steinberg_PRB11,kim2012},
%\\{\bf [cite relevant experiments]}\\
suggest that the theoretical method presented is very effective
to characterize the transport properties of the surfaces of 3DTIs
especially due to its ability to take into account the effects
due to the disorder-induced carrier density inhomogeneities. Future improvement of the theory could include a more accurate surface band structure and the effects of the bulk bands, but we do not expect these details to affect our qualitative results at low surface doping densities because our theory includes the most important resistive processes contributing to the surface transport in 3DTIs.

%Our theoretical results agree semiquantitatively with the recent experimental results
%on thin films of ${\rm Bi_2Se_3}$
%\\{\bf [cite relevant experiments]}\\
%in which the transport of the surface of 3DTIs was studied.

%as is the impurity density, $\nimp\approx 10^{13}{\rm cm}^{-2}$ in 3DTIs
%versus $\nimp\approx 10^{11}{\rm cm}^{-2}$ in graphene

%-----------------------------

\begin{acknowledgments}

Q. L. acknowledges helpful discussions with Euyheon Hwang. This work is supported by ONR-MURI, LPS-CMTC and NRI-SWAN.
ER acknowledges support from the Jeffress Memorial Trust, Grant No. J-1033, and the faculty research grant program from
the College of William \& Mary.

\end{acknowledgments}

%=================

%Merlin.mbs v4.21 2009-07-09.

%Merlin.mbs v4.21 2009-07-09.
%

%xxxxxxxxxxxxxxxxxxxx

\end{document}